\newif\ifanon
\newif\ifsubmission
\submissiontrue

\ifsubmission
  \documentclass[runningheads,orivec]{llncs}
  \providecommand{\keywords}[1]{\par\addvspace\baselineskip
  \noindent{\bf Keywords:}\enspace\ignorespaces#1}%
  \date{}
  \anontrue
    \author{Hamza Jeljeli}
  \institute{CARAMEL project-team, LORIA, INRIA / CNRS / Universit\'e de Lorraine,\\
           Campus Scientifique, BP 239,
           54506 Vand\oe{}uvre-l\`es-Nancy Cedex, France\\
           \url{Hamza.Jeljeli@loria.fr}
  }
  \makeatletter
  \renewcommand\subsubsection{\@startsection{subsubsection}{3}{\z@}%
    {-18\p@ \@plus -4\p@ \@minus -4\p@}%
    {-0.5em \@plus -0.22em \@minus -0.1em}%
    {\normalfont\normalsize\bfseries\boldmath}}
  \makeatother
\else
  \documentclass[runningheads,orivec]{llncs}
  \anonfalse
\fi

\usepackage[utf8]{inputenc}
\usepackage[english]{babel}
\usepackage[T1]{fontenc}
\usepackage{amsmath,amssymb}
\usepackage{hyperref}
\usepackage{multirow}
\usepackage{url}
\usepackage[ruled,lined,noend]{algorithm2e}
\usepackage{pgf}
\usepackage{wrapfig,lipsum}

\usepackage{tikz} 
\usetikzlibrary{arrows}
\usetikzlibrary{snakes}
\usepackage{pgffor,xifthen}
\usepackage{pgfplots}
\usetikzlibrary{patterns}
\usepackage{bm}

\usepackage{listings}
\usepackage{multirow}

\usepackage{tikz} 
\usetikzlibrary{arrows}
\usepgflibrary{arrows}
\usetikzlibrary{matrix}
\usetikzlibrary{decorations.pathreplacing}
\usepackage{tkz-tab}
\usepackage{caption}
\usepackage{latexsym}
\usetikzlibrary{shapes.geometric}
\usepackage{setspace}
\usepackage{subscript}

\usepackage{pgfplots}

\usepackage{subcaption}

\usepackage{array}
\newcolumntype{C}[1]{>{\centering\arraybackslash}p{#1}}

\usepackage{graphicx}

\definecolor{Brown}{cmyk}{0,0.81,1,0.60}
\definecolor{OliveGreen}{cmyk}{0.64,0,0.95,0.40}
\definecolor{CadetBlue}{cmyk}{0.82,0.57,0.13,0}
\definecolor{lightlightgray}{gray}{0.92}

\definecolor{violet}{rgb}{0.5,0,0.5}

\title{Accelerating Iterative SpMV for the Discrete Logarithm Problem Using GPUs}
\titlerunning{Accelerating Iterative SpMV for the DL Problem Using GPUs}

\ifanon\else
\author{Hamza Jeljeli}
\authorrunning{H. Jeljeli}
\institute{CARAMEL project-team, LORIA, INRIA / CNRS / Université de Lorraine,\\
           Campus Scientifique, BP 239,
           54506 Vandœuvre-lès-Nancy Cedex, France\\
           \url{Hamza.Jeljeli@loria.fr}}
\fi

\begin{document}

\maketitle

\begin{abstract}
In the context of cryptanalysis, computing discrete logarithms in large cyclic groups using index-calculus-based methods, such as the number field sieve or the function field sieve, requires solving large sparse systems of linear equations modulo the group order. Most of the fast algorithms used to solve such systems --- e.g., the conjugate gradient or the Lanczos and Wiedemann algorithms --- iterate a product of the corresponding sparse matrix with a vector (SpMV). This central operation can be accelerated on GPUs using specific computing models and addressing patterns, which increase the arithmetic intensity while reducing irregular memory accesses. In this work, we investigate the implementation of SpMV kernels on NVIDIA GPUs, for several representations of the sparse matrix in memory. We explore the use of Residue Number System (RNS) arithmetic to accelerate modular operations. We target linear systems arising when attacking the discrete logarithm problem on groups of size 100 to 1000 bits, which includes the relevant range for current cryptanalytic computations. The proposed SpMV implementation contributed to solving the discrete logarithm problem in GF($2^{619}$) and GF($2^{809}$) using the FFS algorithm.
\end{abstract}


\vspace*{-0.25cm}
\section{Introduction}
\vspace*{-0.25cm}

The security of many cryptographic protocols used for authentication, key exchange, encryption, or signature, depends on the difficulty of solving the discrete logarithm problem (DLP) in a given cyclic group~\cite{ODLY84}. For instance, we can rely on the hardness of the DLP in a multiplicative subgroup of a finite field. There are algorithms, such as Pollard-rho~\cite{POLL75} or Baby-Step/Giant-Step~\cite{SHAN71} that solve the problem in time exponential in the subgroup size. Another family of methods, known as \textit{Index-calculus} methods~\cite{ADLE79} propose to solve it in time sub-exponential or quasi-polynomial in the finite field size. These algorithms require in their linear algebra step the resolution of large sparse systems of linear equations modulo the group order~\cite{LAMA90}. In cryptographic applications, the group order $\ell$ is of size 100 to 1000 bits. The number of rows and columns of the corresponding matrices is in the order of hundreds of thousands to millions, with only hundreds or fewer non-zero elements per row. This linear algebra step is a serious limiting factor in such algorithms. For example, it was reported in~\cite{HAYA12} that the linear algebra step of the Function Field Sieve (FFS) implementation to solve the DLP over GF($3^{6\times97}$) took 80.1 days on 252 CPU cores, which represents 54\% of the total time.
 
To solve such systems, ordinary Gaussian elimination is inefficient. While some elimination strategies aiming at keeping the matrix as sparse as possible can be used to reduce the input system somewhat, actual solving calls for the use of other techniques (Lanczos algorithm~\cite{LANC52}, Wiedemann algorithm~\cite{WIED86}) that take advantage of the sparsity of the matrix~\cite{POME92}. For the Lanczos algorithm, the Wiedemann algorithm and their block variants, the iterative sparse-matrix--vector product is the most time-consuming operation. For this reason, we investigate accelerating this operation on GPUs.

The paper is organized as follows. Section~\ref{Background} presents the background related to the hardware and the context. Section~\ref{RNS} discusses the arithmetic aspects of our implementation. We present several matrix formats and their corresponding implementations in Sections~\ref{SpM formats}. We discuss in Section~\ref{sec::adaptation large integers} how to adapt these implementations over large fields. We compare the results of different implementations run on NVIDIA GPUs in Section~\ref{results}, and present optimizations based on hardware considerations in Section~\ref{optimizations}. Section~\ref{soft} discusses our reference software implementation.  
 
\vspace*{-0.25cm}
\section{Background}
\label{Background}
\vspace*{-0.25cm}

\subsection{GPUs and the CUDA programming model}
\vspace*{-0.25cm}

CUDA is the hardware and software architecture that enables NVIDIA GPUs to execute programs written in C, C++, OpenCL and other languages \cite{CUDA42}.

A CUDA program instantiates a \textit{host} code running on the CPU and a \textit{kernel} code running on the GPU. The kernel code runs according to the Single Program Multiple Threads (SPMT) execution model across a set of parallel threads. The threads are executed in groups of 32, called \textit{warps}. 
If one or more threads have a different execution path, execution divergence occurs. The different paths will then be serialized, negatively impacting the performance. 

The threads are further organized into thread \textit{blocks} and \textit{grids} of thread blocks:
\begin{itemize}
\vspace*{-0.25cm}

\item A thread executes an instance of the kernel. It has a unique thread ID within its thread block, along with registers and private memory.
\item A thread block is a set of threads executing the same kernel which can share data through \textit{shared memory} and perform barrier synchronization which ensures that all threads within that block reach the same instruction before continuing. It has a unique block ID within its grid.
\item A grid is an array of thread blocks executing the same kernel. All the threads of the grid can also read inputs, and write results to \textit{global memory}. 
\end{itemize}

At the hardware level, the blocks are distributed on an array of multi-core \textit{Streaming Multiprocessors} (SMs). Each SM schedules and launches the threads in groups of warps. Recent NVIDIA GPUs of family name ``\textit{Kepler}'' allow for up to 64 active warps per SM. The ratio of active warps to the maximum supported is called \textit{occupancy}. Maximizing the occupancy is important, as it helps to hide the memory latency. One should therefore pay attention to the usage of shared memory and registers in order to maximize occupancy.

Another important performance consideration in programming for the CUDA architecture is \textit{coalescing} global memory accesses. To understand this requirement, global memory should be viewed in terms of aligned segments of 32 words of 32 bits each. Memory requests are serviced for one warp at a time. If the warp requests hit exactly one segment, the access is \textit{fully coalesced} and there will be only one memory transaction performed. If the warp accesses scattered locations, the accesses are \textit{uncoalesced} and there will be as many transactions as the number of hit segments. Consequently, a kernel should use a coalescing-friendly pattern for greater memory efficiency.      
 
Despite their high arithmetic intensity and their large memory bandwidth, GPUs provide small caches. In fact, Kepler GPUs provide the following levels of cache:
\begin{itemize}
\vspace*{-0.25cm}

\item 1536-kB \textit{L2-cache} per GPU.
\item 16-kB \textit{L1-cache} (per SM). It can be extended to 48~kB, but this decreases shared memory from 48~kB to 16~kB.
\item A \textit{texture cache}: an on-chip cache for the read-only \textit{texture memory}. It can accelerate memory accesses when neighboring threads read from nearby addresses.
\end{itemize}  

\vspace*{-0.5cm}

\subsection{Sparse-Matrix--Vector product on GPUs}
\vspace*{-0.25cm}

Sparse-matrix computations pose some difficulties on GPUs, such as irregular memory accesses, load balancing and low cache efficiency. Several papers have focused on choosing suitable matrix formats and appropriate kernels to overcome the irregularity of the sparse matrix \cite{BELL08,VAZQ09}. These works have explored implementing efficiently SpMV over real numbers. Schmidt et al. \cite{SHMI11} proposed an optimized matrix format to accelerate exact SpMV over GF(2), that can be used in the linear algebra step of the Number Field Sieve (NFS) for integer factorization~\cite{STAC08}. Boyer et al. \cite{BOYE10} have adapted SpMV kernels over small finite fields and rings $\mathbb{Z}/m\mathbb{Z}$, where they used double-precision floating-point numbers to represent ring elements. In our context, since the order of the considered finite ring is large (hundreds of bits), specific computing models and addressing models should be used.   

In this work, we have a prime $\ell$, along with an $N$-by-$N$ sparse matrix $A$ defined over $\mathbb{Z}$, and we want to solve the linear system $Aw=0$ over $\mathbb{Z}/\ell\mathbb{Z}$. A feature of the index calculus context that we consider here, is that $A$ contains small values (e.g. 32-bit integers). In fact, around $90\%$ of the non-zero coefficients are $\pm1$.

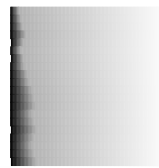
\begin{wrapfigure}[4]{r}{5cm}
\vspace*{-1cm}
\centering
\begin{tikzpicture}[scale=0.6]

\begin{pgfscope}

\pgfdeclarehorizontalshading{myshadingD}
{5.5pt}{rgb(0pt)=(0,0,0); rgb(5pt)=(0.7,0.7,0.7); rgb(100pt)=(1,1,1)}
\pgftext[at=\pgfpoint{0cm}{0cm}] {\pgfuseshading{myshadingD}}

\pgfdeclarehorizontalshading{myshadingD}
{5.5pt}{rgb(0pt)=(0,0,0); rgb(6pt)=(0.7,0.7,0.7 ); rgb(100pt)=(1,1,1)}
\pgftext[at=\pgfpoint{0cm}{-5pt}] {\pgfuseshading{myshadingD}}

\pgfdeclarehorizontalshading{myshadingD}
{5.5pt}{rgb(0pt)=(0,0,0); rgb(7pt)=(0.7,0.7,0.7 ); rgb(100pt)=(1,1,1)}
\pgftext[at=\pgfpoint{0cm}{-10pt}] {\pgfuseshading{myshadingD}}

\pgfdeclarehorizontalshading{myshadingD}
{5.5pt}{rgb(0pt)=(0,0,0); rgb(9pt)=(0.7,0.7,0.7 ); rgb(100pt)=(1,1,1)}
\pgftext[at=\pgfpoint{0cm}{-15pt}] {\pgfuseshading{myshadingD}}

\pgfdeclarehorizontalshading{myshadingD}
{5.5pt}{rgb(0pt)=(0,0,0); rgb(7.5pt)=(0.7,0.7,0.7 ); rgb(100pt)=(1,1,1)}
\pgftext[at=\pgfpoint{0cm}{-20pt}] {\pgfuseshading{myshadingD}}

\pgfdeclarehorizontalshading{myshadingD}
{5.5pt}{rgb(0pt)=(0,0,0); rgb(6pt)=(0.7,0.7,0.7 ); rgb(100pt)=(1,1,1)}
\pgftext[at=\pgfpoint{0cm}{-25pt}] {\pgfuseshading{myshadingD}}

\pgfdeclarehorizontalshading{myshadingD}
{5.5pt}{rgb(0pt)=(0,0,0); rgb(7.5pt)=(0.7,0.7,0.7 ); rgb(100pt)=(1,1,1)}
\pgftext[at=\pgfpoint{0cm}{-30pt}] {\pgfuseshading{myshadingD}}

\pgfdeclarehorizontalshading{myshadingD}
{5.5pt}{rgb(0pt)=(0,0,0); rgb(8pt)=(0.7,0.7,0.7 ); rgb(100pt)=(1,1,1)}
\pgftext[at=\pgfpoint{0cm}{-35pt}] {\pgfuseshading{myshadingD}}

\pgfdeclarehorizontalshading{myshadingD}
{5.5pt}{rgb(0pt)=(0,0,0); rgb(10pt)=(0.7,0.7,0.7 ); rgb(100pt)=(1,1,1)}
\pgftext[at=\pgfpoint{0cm}{-40pt}] {\pgfuseshading{myshadingD}}

\pgfdeclarehorizontalshading{myshadingD}
{5.5pt}{rgb(0pt)=(0,0,0); rgb(11pt)=(0.7,0.7,0.7 ); rgb(100pt)=(1,1,1)}
\pgftext[at=\pgfpoint{0cm}{-45pt}] {\pgfuseshading{myshadingD}}

\pgfdeclarehorizontalshading{myshadingD}
{5.5pt}{rgb(0pt)=(0,0,0); rgb(13pt)=(0.7,0.7,0.7 ); rgb(100pt)=(1,1,1)}
\pgftext[at=\pgfpoint{0cm}{-50pt}] {\pgfuseshading{myshadingD}}

\pgfdeclarehorizontalshading{myshadingD}
{5.5pt}{rgb(0pt)=(0,0,0); rgb(14pt)=(0.7,0.7,0.7 ); rgb(100pt)=(1,1,1)}
\pgftext[at=\pgfpoint{0cm}{-55pt}] {\pgfuseshading{myshadingD}}

\pgfdeclarehorizontalshading{myshadingD}
{5.5pt}{rgb(0pt)=(0,0,0); rgb(15pt)=(0.7,0.7,0.7 ); rgb(100pt)=(1,1,1)}
\pgftext[at=\pgfpoint{0cm}{-60pt}] {\pgfuseshading{myshadingD}}

\pgfdeclarehorizontalshading{myshadingD}
{5.5pt}{rgb(0pt)=(0,0,0); rgb(13pt)=(0.7,0.7,0.7 ); rgb(100pt)=(1,1,1)}
\pgftext[at=\pgfpoint{0cm}{-65pt}] {\pgfuseshading{myshadingD}}

\pgfdeclarehorizontalshading{myshadingD}
{5.5pt}{rgb(0pt)=(0,0,0); rgb(14pt)=(0.7,0.7,0.7 ); rgb(100pt)=(1,1,1)}
\pgftext[at=\pgfpoint{0cm}{-70pt}] {\pgfuseshading{myshadingD}}

\pgfdeclarehorizontalshading{myshadingD}
{5.5pt}{rgb(0pt)=(0,0,0); rgb(15.5pt)=(0.7,0.7,0.7 ); rgb(100pt)=(1,1,1)}
\pgftext[at=\pgfpoint{0cm}{-75pt}] {\pgfuseshading{myshadingD}}

\pgfdeclarehorizontalshading{myshadingD}
{5.5pt}{rgb(0pt)=(0,0,0); rgb(14pt)=(0.7,0.7,0.7 ); rgb(100pt)=(1,1,1)}
\pgftext[at=\pgfpoint{0cm}{-80pt}] {\pgfuseshading{myshadingD}}

\pgfdeclarehorizontalshading{myshadingD}
{5.5pt}{rgb(0pt)=(0,0,0); rgb(13pt)=(0.7,0.7,0.7 ); rgb(100pt)=(1,1,1)}
\pgftext[at=\pgfpoint{0cm}{-85pt}] {\pgfuseshading{myshadingD}}

\pgfdeclarehorizontalshading{myshadingD}
{5.5pt}{rgb(0pt)=(0,0,0); rgb(14pt)=(0.7,0.7,0.7 ); rgb(100pt)=(1,1,1)}
\pgftext[at=\pgfpoint{0cm}{-90pt}] {\pgfuseshading{myshadingD}}

\pgfdeclarehorizontalshading{myshadingD}
{5.5pt}{rgb(0pt)=(0,0,0); rgb(15.5pt)=(0.7,0.7,0.7 ); rgb(100pt)=(1,1,1)}
\pgftext[at=\pgfpoint{0cm}{-95pt}] {\pgfuseshading{myshadingD}}

\end{pgfscope}

\end{tikzpicture}

\caption{Distribution of non-zero elements in an FFS matrix}
\label{FFS matrix}
\end{wrapfigure}

The very first columns of $A$ are relatively dense, then the column density decreases gradually. The row density does not change significantly. We denote by $n_{\mathrm{NZ}}$ the number of non-zero elements in $A$. See Figure~\ref{FFS matrix} for a typical density plot of a matrix arising in an FFS computation.

We will use the Wiedemann algorithm as a solver. This algorithm iterates a very large number of matrix-vector products of the form $v \leftarrow Au$, where $u$ and $v$ are dense $N$-coordinate vectors. The major part of this work deals with how to accelerate this product.

In order to carry out this product, we compute the dot product between each row of $A$ and the vector $u$. The basic operation is of the form $x \leftarrow (x + \lambda y) \bmod \ell$, where $\lambda$ is a non-zero coefficient of $A$, and $x$ and $y$ are coordinates of the vectors $v$ and $u$, respectively. To minimize the number of costly reductions modulo $\ell$, we accumulate computations, and postpone the final modular reduction of the result as late as possible. When iterating many products (computations of the form $A^i u$), we can further accumulate several SpMVs before reducing modulo $\ell$, as long as the intermediate results do not exceed the largest representable integer. As far as arithmetic over $\mathbb{Z}/\ell\mathbb{Z}$ is concerned, we chose to use the Residue Number System, which appears to be more suited to the fine grained parallelism inherent to the SPMT computing model than the usual multi-precision representation of large integers. A comparison of the two representations is given in Subsection~\ref{RNS vs MP}.

\vspace*{-0.25cm}
\section{Residue Number System and Modular Arithmetic}
\label{RNS}

\vspace*{-0.25cm}

\subsection{A brief reminder on RNS}
\vspace*{-0.25cm}

The Residue Number System (RNS) is based on the Chinese Remainder Theorem (CRT). Let $\mathcal{B}=(p_1,p_2,\dots,p_n)$ be a set of mutually coprime integers, which we call an \textit{RNS-basis}. We define $P$ as the product of all the $p_i$'s. The RNS uses the fact that any integer $x$ within $[0, P-1]$ can be uniquely represented by the list $(x_1,x_2,\dots,x_n)$, where each $x_i$ is the residue of $x$ modulo $p_i$, which we write as $x_i = |x|_{p_i}$. 

If $x$ and $y$ are given in their RNS representations $\vec{x}=(x_1,\dots,x_n)$ and $\vec{y}=(y_1,\dots,y_n)$, according to $\mathcal{B}$, and such that $x,y<P$, RNS addition and multiplication are realized by modular addition and multiplication on each component:
\begin{equation*}
\vec{x}\vec{+}\vec{y} = (|x_1+y_1|_{p_1},\dots,|{x_n}+{y_n}|_{p_n}),\,\,\,\, \vec{x}\vec{\times}\vec{y} = (|x_1\times y_1|_{p_1},\dots, |x_n \times y_n|_{p_n})
\end{equation*}
The result (e.g., $x+y$) should belong to the interval $[0,P-1]$ if we want to obtain a valid RNS representation. Otherwise, it will be reduced modulo $P$. Unlike addition or multiplication, other operations such as comparison or modular reduction are more subtle in RNS.
 
We can convert back an RNS vector to the integer form by using the CRT formula:

\begin{equation}
x = \displaystyle \left|\sum_{i=1}^{n}x_i \cdot \left|{P_i^{-1}}\right|_{p_i}\cdot{P_i} \right|_ {P} \mathrm{, where }\; P_i \triangleq P/p_i.
\nonumber
\end{equation}

This number system is particularly interesting for arithmetic over large integers, since it distributes the computation over several small residues. In other words, the computation units that will work on the residues are independent and need no synchronization nor communication, as there is no carry propagation \cite{TANA67,TAYL84}.

\vspace*{-0.25cm}

\subsection{RNS reduction modulo $\ell$}
\vspace*{-0.25cm}
\label{modular reduction}

%

In the chosen RNS representation, $(P-1)$ is the largest representable integer. So in the case of repeated SpMVs over $\mathbb{Z}/\ell\mathbb{Z}$, we can accumulate at most $\log(\frac{P-1}{\ell-1})/\log(r)$ matrix--vector products before having to reduce modulo $\ell$, where $r$ corresponds to the largest row norm (defined as the sum of the absolute values of its elements) in the matrix. To reduce the vector $v$ modulo $\ell$, we use the method introduced by Bernstein in \cite{BERN95}, which allows us to perform the reduction without having to convert the vector back to the integer form.

We assume that the RNS-basis $\mathcal{B}$ contains $n$ moduli $p_1,\dots,p_n$ of $k$ bits each. We impose that the $p_i$'s are close to $2^k$. The reasons will be detailed in the following subsection. We want to reduce modulo $\ell$ an RNS vector $(x_1,\dots,x_n)$. We start from the CRT reconstruction:  
 $x = \displaystyle \left|\sum_{i=1}^{n}{\gamma_i P_i} \right|_ {P}$, where we have defined $\gamma_i \triangleq {\left|{x_iP_i}^{-1}\right|_{p_i}}$. Let us also define the integer $\alpha$ as follows 

\begin{equation}
\alpha = \left \lfloor \displaystyle \sum_{i=1}^{n} { \displaystyle\frac{\gamma_i P_i}{P}} \right \rfloor = \left \lfloor \displaystyle \sum_{i=1}^{n} { \displaystyle\frac{\gamma_i}{p_i}} \right \rfloor.
\end{equation}

The vector $x$ can then be written as $\displaystyle\sum_{i=1}^{n}{\gamma_i{P_i}} - \alpha P$ and, since $\gamma_i < p_i$, we have $ 0 \leq \alpha < n$.
 
Now, if we assume that $\alpha$ is known, we define $\displaystyle {z \triangleq \sum_{i=1}^{n}{\gamma_i\left|P_i\right|_{\ell}} - \left|\alpha P\right|_{\ell}}$. We can easily check that $z$ is congruent to $x$ modulo $\ell$ and lies in the interval $[0,\ell \displaystyle\sum_{i=1}^{n}{p_i}[$.

What remains to be done is to determine $\alpha$.
Since $p_i \approx 2^k$, we approximate the quotient $\gamma_i / p_i$ using only the $s$ most significant bits of $\gamma_i / 2^k$. Hence, an estimate for $\alpha$ is proposed as

\begin{equation}
\hat{\alpha}\triangleq \displaystyle \left\lfloor { \sum_{i=1}^{n} { \displaystyle \frac{\left \lfloor \displaystyle \frac{\gamma_i}{2^{k-s}} \right \rfloor} {2^s} + \Delta}}\right\rfloor,
\end{equation}
where $s$ is an integer parameter in $[1,k]$ and $\Delta$ an error correcting term in $]0,1[$. 

Bernstein states in \cite{BERN95} that if $0\leq x<(1-\Delta) P$ and $(\epsilon + \delta) \leq \Delta < 1$ where $\epsilon \triangleq \displaystyle \sum_{i=1}^{n} \displaystyle \frac{c_i}{2^k}$ and $ \delta \triangleq n \displaystyle \frac{2^{k-s}-1}{2^k} $, then $\alpha = \hat{\alpha}$.  

Once $\alpha$ is determined, we are able to perform an RNS computation of $z$. Algorithm~\ref{RNS modular reduction} summarizes the steps of the computation. 
  
\begin{algorithm}[H]
  \LinesNumbered
  \SetAlgoLined
  \SetKwInOut{Input}{Precomputed}
  \SetKwInOut{Output}{Output}
  \DontPrintSemicolon 
  \Input{
  Vector $(\left|{P_j}^{-1}\right|_{p_j})$ for $j \in \{1, \dots, n\}$\\
  $\,\,\,\,\,\,\,\,\,\,\,\,\,\,\,\,\,\,\,$ Table of RNS vectors of $\left|P_i\right|_\ell$ for $i \in \{1, \dots, n\}$\\
  $\,\,\,\,\,\,\,\,\,\,\,\,\,\,\,\,\,\,\,$ Table of RNS vectors of $\left|\alpha P\right|_\ell$ for $\alpha \in \{1, \dots, n-1\}$}
  \SetKwInOut{Input}{Input}
  \Input{RNS vector of $x$, with $0 \leq x < (1-\Delta) P$}
  \Output{RNS vector of $z\equiv x \pmod{{\ell}}$, $z<\ell \displaystyle\sum_{i=1}^{n}{p_i}$}
  \ForEach{thread j}{
   $\gamma_j \gets \left|x_j \times {\left|{P_j}^{-1}\right|_{p_j}}\right|_{p_j}$ \tcc*[r]{1 RNS product}
}
  Broadcast of the $\gamma_j$'s by all the threads\\
  \ForEach{thread j}{
  \small $z_j \gets \left|\displaystyle \sum_{i=1}^{n}{\gamma_i \times \left|{\left|{P_i}\right|_{\ell}}\right|_{p_j}  }\right|_{p_j}$ \tcc*[r]{$(n-1)$ RNS sums \& $n$ RNS products}
  $\alpha \gets \displaystyle \left\lfloor { \sum_{i=1}^{n} { \frac{\left \lfloor \displaystyle \frac{\gamma_i}{2^{k-s}} \right \rfloor} {2^s} + \Delta}}\right\rfloor$ \tcc*[r]{sum of $n$ $s$-bit terms}
    $z_j \gets  \left|z_j - {\left|{\left|\alpha P\right|_{\ell}}\right|_{p_j}}\right|_{p_j}$ \tcc*[r]{1 RNS subtraction}
  }  
  \caption{Approximate RNS modular reduction}
  \label{RNS modular reduction}
\end{algorithm}

All the operations can be evaluated in parallel on the residues, except for step 3, where a broadcast of all the $\gamma_j$'s is needed. Even if the obtained result $z$ is not the exact reduction of $x$, it is bounded by $n2^k{\ell}$. Thus, we guarantee that the intermediate results of the SpMV computation do not exceed a certain bound less than $P$. Notice that this RNS reduction algorithm imposes that $P$ be one modulus ($k$ bits) larger than implied by the earlier condition $\ell<P$.

In conclusion, $P$ is chosen, such that $r \times n2^{k}\ell < (1-\Delta)P$, with $r$ is the largest row norm of the matrix.

\vspace*{-0.25cm}

\subsection{Modular reduction modulo $p_j$}
\vspace*{-0.25cm}

The basic RNS operation is $z_j \leftarrow (x_j + \lambda \times y_j)\bmod p_j$, where $0 \leq x_j,y_j,z_j < p_j$ are RNS residues and $\lambda$ is a positive element of the matrix. So, it consists of an AddMul (multiplication, then an addition) followed by a reduction modulo $p_j$. To speed up the reduction modulo $p_j$, the moduli are chosen of the pseudo-Mersenne form $2^{k}-c_j$, with $c_j$ as small as possible. 

In fact, let us define $t_j \triangleq x_j + \lambda \times y_j$ as the intermediate result before the modular reduction. $t_j$ can be written as
\begin{equation}
t_j = t_{jL} + 2^k \times t_{jH}, \textrm{where } t_{jL} \triangleq t_j \bmod 2^{k}, t_{jH} \triangleq t_j / 2^{k}. 
\end{equation}
Since $2^{k} \equiv c_j \pmod {p_j}$, we have
$t_j \equiv t_{jL} + t_{jH} \times c_j \pmod {p_j}$.
So, we compute $t_j \gets t_{jL} + t_{jH} \times c_j$, then we have to consider two cases:
\begin{itemize}
\item if $t_j < 2^k$, we have ``almost" reduced $(x_j + \lambda \times y_j)$ modulo $p_j$, since the result lies in $[0,2^k[$, not in $[0,p_j[$;
\item else we have reduced $t_j$ by approximately $k$ bits. Thus, we repeat the previous procedure with $t_j \leftarrow t_{jL} + c_j \times t_{jH} $, which then satisfies $t_j<2^k$.
\end{itemize} 

The output lies in $[0,2^k-1]$, so we propose to relax the condition on both input and output: $x_j, z_j \in [0,2^{k}-1]$. With this approach, the reduction can be done in a small number of additions and products.

\vspace*{-0.25cm}

\subsection{Possible RNS Mappings on GPU/CPU}
\vspace*{-0.25cm}

We represent the finite ring $\mathbb{Z}/\ell\mathbb{Z}$ as the integer interval $[0,\ell-1]$. Each element is stored in its RNS form. On GPU, we opted for 64-bit moduli (i.e.~$k=64$), for performance considerations. Even that floating point instructions have higher throughput, integer instructions gave better performances, because with floating point arithmetic, only the mantissa is used and the algorithms are more complex than with integer arithmetic. 
We use the PTX (\textit{parallel thread execution}) pseudo-assembly language for CUDA \cite{PTX30} to implement the RNS operations.
\medskip

On CPU, we implemented three versions based on:
\vspace*{-0.25cm}
\begin{itemize}
\item MMX instruction set: we map an RNS residue to an unsigned 64-bit integer.
\item Streaming SIMD Extensions (SSE2) set: a 128-bit XMM register holds two residues, so the processor can process two residues simultaneously.
\item Advanced Vector Extensions (AVX2) set: we use the 256-bit YMM register to hold four residues. 
\end{itemize} 

\vspace*{-0.5cm}
\section{Sparse Matrix Storage Formats}

In this section, we assume that the elements of the matrix, as well as the elements of the vectors $u$ and $v$ are in a field $K$ (reals, finite fields, etc.). For each format, we will discuss how to perform the matrix-vector product. We will give a pseudo-code for the format CSR. Figures that illustrate the other formats and their corresponding Pseudo-code can be found in Appendix~\ref{Appendix:: kernels}.

The matrix and vectors are put in \textit{global memory}, since their sizes are important. Temporary results are stored in registers. The \textit{shared memory} is used when partial results of different threads are combined. Arithmetic operations are performed in registers and denoted in the pseudo-code by the function \texttt{addmul()}.

\label{SpM formats}

\vspace*{-0.25cm}
\subsection*{Coordinate (COO)}
\vspace*{-0.25cm}

The format COO consists of three arrays \texttt{row\_id}, \texttt{col\_id} and \texttt{data} of $n_{\mathrm{NZ}}$ elements. The row index, column index and the value are explicitly stored to specify a non-zero matrix coefficient. In this work, we propose to sort the matrix coefficients by their row index.

A typical way to work with the COO format on GPU is to assign one thread to each non-zero matrix coefficient. This implies that different threads from different warps will process a same row. 
Each thread computes its partial result, then performs a segmented reduction \cite{BLEL93,SENG07} to sum the partial results of the other threads belonging to the same warp and spanning the same row. We followed the scheme proposed by the library CUSP~\cite{CUSP}, which performs the segmented reduction in shared memory, using the row indices as segment descriptors. Each warp iterates over its interval, processing 32 coefficients at a time. If a spanned row is fully processed, its result is written to $v$, otherwise, the row index and the partial dot product are stored in temporary arrays. Then, a second kernel performs the combination of the per-warp results.

The main drawbacks of the COO kernel are the cost of the combination of partial results and excessive usage of \textit{global memory}. Its advantage is that the workload distribution is balanced across warps, as they iterate over a constant length interval.   

\vspace*{-0.25cm}

\subsection*{Compressed Sparse Row (CSR)} 
\vspace*{-0.25cm}

The CSR format stores the column indices and the values of the non-zero elements of $A$ into two arrays of $n_{\mathrm{NZ}}$ elements: \texttt{id} and \texttt{data}. A third array of pointers, \texttt{ptr}, of length $N+1$, is used to indicate the beginning and the end of each row. Non-zero coefficients are sorted by their row index. The CSR format eliminates the explicit storage of the row index, and is convenient for a direct access to the matrix, since \texttt{ptr} indicates where each row starts and ends in the other two ordered arrays.

\begin{figure}[]

\vspace*{-1cm}

\centering

\begin{tabular}{C{.36\textwidth}C{.64\textwidth}}

\begin{subfigure}[]{0.35\textwidth}
	\centering
	\begin{tikzpicture}[scale=0.8]
		\draw [transparent](-2,-2) rectangle (2,2);
		\tikzstyle{every right delimiter}=[xshift=-1.5ex]
		\tikzstyle{every left delimiter}=[xshift=1.5ex]
		\pgflowlevelsynccm
		\matrix [matrix of math nodes,left delimiter=(,right delimiter=)] at (0,0)
		{
		0 & \color{red} a_{01} & 0 & \color{red} a_{03} & 0 & 0 \\
		0 & \color{blue} a_{11} & 0 & 0 & \color{blue} a_{14} & \color{blue} a_{15} \\
		\color{orange} a_{20} & 0 & \color{orange} a_{22} & \color{orange} a_{23} & 0 & 0 \\
		0 & a_{31} & 0 & 0 & a_{34} & 0 \\
		0 & a_{41} & a_{42} & 0 & 0 & a_{45} \\
		0 & 0 & a_{52} & 0 & 0 & a_{55} \\
		};
	\end{tikzpicture}
	\caption{Sparse matrix $A$}
\end{subfigure}

& 

\begin{subfigure}[]{0.63\textwidth}

	\begin{tikzpicture}[scale=0.8]
		\draw [transparent](-5.15,0.75) rectangle (3.35,-2.5);
			\pgflowlevelsynccm
			\node at (-4.5,0){\texttt{data = }};
			\tikzstyle{every right delimiter}=[xshift=-1.5ex]
			\tikzstyle{every left delimiter}=[xshift=1.5ex]
			\matrix [matrix of math nodes,left delimiter={[},right delimiter={]}] at (-0.25,0)
			{
			\color{red} a_{01} & \color{red} a_{03} & \color{blue} a_{11} & \color{blue} a_{14} & \color{blue} a_{15} & \color{orange} a_{20} & \color{orange} a_{22} & \color{orange} a_{23} & \dots\\
			};
			
			\node at (-4.25,-1){\texttt{id = }};
			\tikzstyle{every right delimiter}=[xshift=-1.5ex]
			\tikzstyle{every left delimiter}=[xshift=1.5ex]
			\matrix [matrix of math nodes,style={nodes={minimum width=2em}},left delimiter={[},right delimiter={]}] at (-0.25,-1)
			{
			\color{red} 1 & \color{red} 3 & \color{blue} 1 & \color{blue} 4 & \color{blue} 5 & \color{orange} 0 & \color{orange} 2 & \color{orange} 3 & \dots\\
			};
			
			\node at (-4.3,-2){\texttt{ptr = }};
			\tikzstyle{every right delimiter}=[xshift=-1.5ex]
			\tikzstyle{every left delimiter}=[xshift=1.5ex]
			\matrix [matrix of math nodes,style={nodes={minimum width=2em}},left delimiter={[},right delimiter={]}] at(-1.675,-2)
			{
			\color{red} 0 & \color{blue} 2 & \color{orange} 5 & 8 &  &  &  &  & \dots\\
			};	
	\end{tikzpicture}
	\caption{CSR representation}	
	
\end{subfigure}

\end{tabular}

\vspace*{-1.25cm}

\end{figure}

\vspace*{-0.25cm}

\subsubsection{Scalar approach (CSR-S)}

To parallelize the product for the CSR format, a simple way is to assign each row to a single thread (\textit{scalar} approach).
For each non-zero coefficient, the thread performs a read from \textit{global memory}, an \texttt{addmul} and a write in \textit{registers}. Final result is written to \textit{global memory}.


\vspace*{-0.25cm}

\subsubsection{Vector approach (CSR-V)}

The \textit{vector} approach consists in assigning a warp to each row of the matrix~\cite{BELL08}. The threads within a warp access neighboring non-zero elements, which makes the warp accesses to \texttt{id} and \texttt{data} contiguous. Each thread computes its partial result in shared memory, then a parallel reduction in shared memory is required to combine the per-thread results (denoted \texttt{reduction\_csr\_v()} in Algorithm~\ref{Kernel::CSR}). No synchronization is needed, since threads belonging to a same warp are implicitly synchronized.




\begin{algorithm}[]
 \SetKwInOut{Input}{Inputs}\SetKwInOut{Output}{Output}
 \SetKwFor{For}{For}{do}{endfor}
 \SetKwFor{While}{While}{do}{endwhile}
 \SetKwFor{If}{If}{then}{else}
 \Input{\small \texttt{data}: array of $n_{NZ}$ elements of $K$, \texttt{id}: array of $n_{NZ}$ positive integers,\\ $\,$ \texttt{ptr}: array of $N$ positive integers and $u$: vector of $N$ elements of $K$.
 }
 \Output{\small $v$: vector of $N$ elements of $K$.}
 \BlankLine
 \small 
 \texttt{sum} $\leftarrow 0$\;
 $j\leftarrow$ \texttt{ptr}\textsubscript{$i$} $+$ \texttt{tid}\tcp*[r]{position of beginning for each thread}
  \While{$j<$ \texttt{\upshape ptr\textsubscript{$i+1$}}}{
   \texttt{sum} $\leftarrow$ \texttt{addmul(sum,data}\textsubscript{$j$}\texttt{,}$u$\textsubscript{\texttt{id}\textsubscript{$j$}}\texttt{)}\;
   $j \leftarrow j + 32$\;
   }
   \texttt{reduction\_csr\_v(sum,tid)}\tcp*[r]{reduction in \textit{shared memory}}
  \If(\tcp*[f]{first thread of the warp writes in \textit{global memory}}){\texttt{\upshape tid}$=0$}{
  	$v_i$ $\leftarrow$ \texttt{sum}\;
  } 
 \caption{\small CSR-V for row $i$ executed by thread of index \texttt{tid} in its warp}
\label{Kernel::CSR}

\end{algorithm}


Compared to COO kernel, the two CSR kernels reduce the usage of \textit{global memory} and simplify the combination of partial results. The CSR kernels suffer from load unbalance, if the rows have widely varying lengths. To improve the load balance, one possibility is to order the rows by their lengths. So, the warps launched simultaneously have almost the same load.

If we compare the two CSR kernels. The threads of CSR-S have non contiguous access to \texttt{data} et \texttt{id}, as they do not work on the same rows. Thus, their memory accesses are not as efficient as the accesses of the CSR-V. However, the CSR-V kernel requires a combination of partial results which increases the use of \textit{registers} and \textit{shared memory} (cf. Subsection~\ref{subsection: comparison kernels}).

\vspace*{-0.25cm}

\subsection*{ELLPACK (ELL)}
\vspace*{-0.25cm}

The ELL format extends the CSR arrays to $N$-by-$K$ arrays, where $K$ corresponds to the maximum number of non-zero coefficients per row. The rows that have less than $K$ non-zero coefficients are padded. Since the padded rows have the same length, only column indices are explicitly stored. This format suffers from the overhead due to the padding when the the percentage of zeros is high. An optimization was proposed by V\'{a}zquez et al. with a format called ELLPACK-R (ELL-R)~\cite{VAZQ09}. This variant adds an array \texttt{len} of length $N$ that indicates the length of each row. Thus, the zeros added by the padding are not considered when performing the matrix-vector product.

The partitioning of the work is done by assigning a thread to a row of the matrix. The kernel takes advantage from the column-major ordering of the elements to improve the accesses on the vector $u$. However, it suffers from thread divergence. 
 
\vspace*{-0.25cm}

\subsection*{Sliced Coordinate (SLCOO)}
\vspace*{-0.25cm}

The SLCOO format was introduced on GPUs by Schmidt et al. for integer factorization, in the particular case of matrices over GF(2) \cite{SHMI11} and was inspired by the CADO-NFS~\cite{CADO} software for CPUs. The aim of this format is to increase the cache hit rate that limits the CSR and COO performance. 
Like COO, the SLCOO representation stores the row indices, column indices and values. However, it divides the matrix into horizontal slices, where the non-zero coefficients of a slice are sorted according to their column index in order to reduce the irregular accesses on source vector $u$, if they had been sorted by their row indices. A fourth array \texttt{ptrSlice} indicates the beginning and end of each slice. We denote this format SLCOO-$\sigma$, where the parameter $\sigma$ is the number of rows in a slice.

For the SLCOO kernel, each warp works on a slice. Since each thread works on more than one row, it needs to have individual storage for its partial per-row results, or to be able to have exclusive access to a common resource. In~\cite{SHMI11}, Schmidt et al. mentionned the two possibilities of either using the \textit{shared memory} or having atomic accesses. While these needs can be fulfilled in~\cite{SHMI11} for the context of linear algebra over GF(2), we will observe in Section~\ref{results} that these constraints hamper the efficiency of the SLCOO in the context of large fields.

\medskip
There are other SpMV formats in the literature, such as DIA (Diagonal) format, that are appropriate only for matrices that satisfy some sparsity patterns, which is not our case.

\vspace*{-0.25cm}
\section{SpMV Kernels over large fields}
\label{sec::adaptation large integers}

In the context of our application, the matrix elements are ``small'' (32 bit integers) and the vectors elements are in $\mathbb{Z}/\ell\mathbb{Z}$. In this section, we study how we adapt the kernels to this context. We assume that an element of $\mathbb{Z}/\ell\mathbb{Z}$ holds in $n$ machine words. Thus, processing a non-zero coefficient $\lambda$ at row $i$ and column $j$ of the matrix implies reading the $n$ words that compose the $j^{th}$ element in the input vector $u$, multiply them by $\lambda$ and adding them to the $n$ words that compose the $i^{th}$ element in the output vector $v$. In the following pseudo-code, we denote the arithmetic operation that applies to a word by the function \texttt{addmul\_word()}. Pseudo-code is therefore given without details regarding the representation system of the numbers and the resulting arithmetic.

\vspace*{-0.25cm}
\paragraph{\bf \textit{Sequential} Scheme}

A first approach would be that each thread processes a coefficient. We would call this scheme \textit{sequential}. To illustrate this scheme, we apply it on the CSR-V kernel.


\begin{algorithm}[H]
 \SetKwInOut{Input}{Inputs}\SetKwInOut{Output}{Output}
 \SetKwFor{For}{For}{do}{endfor}
 \SetKwFor{While}{While}{do}{endwhile}
 \SetKwFor{If}{If}{then}{else}
 \Input{\small \texttt{data}: array of $n_{NZ}$ signed integers,
 \texttt{id}: array of $n_{NZ}$ positive integers,\\ $\,$ \texttt{ptr}: array of $N$ positive integers,
 $u$: vector of $N \times n$ machine words.
 }
 \Output{\small $v$: vector of $N \times n$ machine words.}
 \small 
 \BlankLine
 Declare array \texttt{sum} $\leftarrow \{0\}$\tcp*[r]{$n$ machine words initialized to $0$}
 $j\leftarrow$ \texttt{ptr}\textsubscript{$i$} $+$ \texttt{tid}\;
  \While{$j<$ \texttt{\upshape ptr\textsubscript{$i+1$}}}{
   \For{$k\leftarrow 0$ \KwTo $n$}{
		\texttt{sum}\textsubscript{$k$} $\leftarrow$ \texttt{addmul\_word(sum}\textsubscript{$k$}\texttt{,data}\textsubscript{$j$}\texttt{,}$u$\textsubscript{\texttt{id}\textsubscript{$j$}$\times n + k$}\texttt{)}\tcp*[r]{\small process $k$\textsuperscript{th} word}
   }
   $j \leftarrow j + 32$\;
   }
   \texttt{reduction\_csr\_v\_seq(sum,tid)}\;
  \If{\texttt{\upshape tid}$=0$}{
  	\For{$k\leftarrow 0$ \KwTo $n$}{
		$v_{i \times n + k}$ $\leftarrow$ \texttt{sum}\textsubscript{$k$}\tcp*[r]{store $k$\textsuperscript{th} word in \textit{global memory}}		
    }
  } 
 \caption{\small CSR-V-seq for row $i$ executed by thread of index \texttt{tid} in its warp}
 \label{algo::csr vector-sequential}
\end{algorithm}


This scheme suffers from several drawbacks. The first one is that the thread processes the $n$ machine words corresponding to the coefficient, i.e. reads and writes the $n$ machine words and makes $n$ arithmetic operations. Thus, the thread consumes more \textit{registers}. The second is that the threads of the same warp access non-contiguous zones of the vectors $u$ and $v$, as their accesses are always spaced by $n$ words.

\vspace*{-0.25cm}
\paragraph{\bf \textit{Parallel} Scheme}

To overcome the limitations of the previous approach, a better scheme would be that a nonzero coefficient is processed by $n$ threads. We refer to the scheme by the \textit{parallel} scheme. Threads of the same warp are organized in $n_{\mathrm{GPS}}$ of $n$ threads, where $ n_{\mathrm{GPS}} \times n$ is closest to 32, the number of threads per warp. Each group is associated to a non-zero matrix coefficient. For example, for n = $5$, we take $n_{\mathrm{GPS}} = 6$, so the first 5 threads process in parallel the 5 words of the 1\textsuperscript{st} source vector element, threads 5 to 9, process the words of the 2\textsuperscript{nd} source vector element, and so on, and we will have two idle threads per warp.


\begin{algorithm}[H]
 \SetKwInOut{Input}{Inputs}\SetKwInOut{Output}{Output}
 \SetKwFor{For}{For}{do}{endfor}
 \SetKwFor{While}{While}{do}{endwhile}
 \SetKwFor{If}{If}{then}{else}
 \Input{\small \texttt{data}: array of $n_{NZ}$ signed integers,
 \texttt{id}: array of $n_{NZ}$ positive integers,
 \\ $\,$ \texttt{ptr}: array of $N$ positive integers,
 $u$: vector of $N \times n$ mots machines.
 }
 \Output{\small $v$: vector of $N \times n$ mots machines.}
 \small 
 \BlankLine
 \texttt{sum} $\leftarrow 0$\tcp*[r]{1 machine word initialized to $0$}
 $j\leftarrow$ \texttt{ptr}\textsubscript{$i$} $+ \lfloor$ \texttt{tid} $/$ $n \rfloor$\tcp*[r]{position of beginning for each thread}
  \While{$j<$ \texttt{\upshape ptr\textsubscript{$i+1$}}}{
   \texttt{sum} $\leftarrow$ \texttt{addmul\_word(sum,}\texttt{data}\textsubscript{$j$}
   \texttt{,}$u$\textsubscript{\texttt{id}\textsubscript{$j$}$\times n$ $+$ \texttt{tid} $\bmod n$}\texttt{)}\tcp*[r]{process 1 word}
   $j \leftarrow j + n_{\mathrm{GPS}}$\;
   }
   \texttt{reduction\_csr\_v\_par(sum,tid)}\;
  \If(\tcp*[f]{first group of the warp writes in \textit{global memory}}){\texttt{\upshape tid}$<n$}{
  	$v$\textsubscript{$i\times n$ $+$ \texttt{tid}} $\leftarrow$ \texttt{sum}\;
  } 
 \caption{\small CSR-V-par for row $i$ executed by thread of index \texttt{tid} in its warp}
 \label{algo::csr vector-parallel}
\end{algorithm}


\medskip
For the other kernels, both schemes are applicable and the \textit{parallel} scheme always performs significantly better than the \textit{sequential} scheme.

\vspace*{-0.25cm}

\section{Comparative Analysis of SpMV Kernels}
\label{results}
In this section, we compare the performances of the kernels that we presented. The objective is to minimize the time of a matrix-vector product. The experiments were run on an NVIDIA GeForce GTX 680 graphics processor (Kepler). Each SpMV kernel was executed 100 times, which is large enough to obtain stable timings.
Our measurements do not include the time spent to copy data between the host and the GPU, since the matrix and vectors do not need to be transferred back and forth between each SpMV iteration. The reduction modulo $\ell$ happens only once every few iterations, which is why the timing of an iteration includes the timing of the reduction modulo $\ell$ kernel multiplied by the frequency of its invocation. The reported measurements are based on NVIDIA developer tools.


\begin{wraptable}[8]{r}{6.25cm}
\vspace*{-1cm}
  \centering
    \begin{tabular}{|l||c|}
      \hline
      \small Size of the matrix ($N$) & \footnotesize 650k $\times$ 650k\\
      \small \#Non-zero coefficients & \footnotesize 65M\\
      \small Max (row norm) &\footnotesize 492\\
      \small Percentage of $\pm1$ & \footnotesize 92.7\%\\
      \small Size of $\ell$ (in bits)& \footnotesize 217\\
      \small Size of $M$ (in bits)& \footnotesize 320\\
      \small Size of $n2^k{\ell}$ (in bits)& \footnotesize 283\\
      \small Frequency of reduction $\bmod \ell$ & \footnotesize 1/4\\      
      \hline
    \end{tabular}
  \caption{ \small{Properties of test matrix}}
  \label{Matrices}
\end{wraptable}

Table~\ref{Matrices} summarizes the considered matrix over $\mathbb{Z}/{\ell}\mathbb{Z}$. The matrix was obtained during the resolution of discrete logarithm problem in the 217-bit prime order subgroup of GF($2^{619}$)$^\times$ using the FFS algorithm. The $\mathbb{Z}/{\ell}\mathbb{Z}$ elements fit in four RNS 64-bit residues. Since, an extra residue is needed for the modular reduction (cf. Subsection~\ref{modular reduction}), the total number of RNS residues is $n=5$.

\vspace*{-0.25cm}
\subsection{Comparison of schemes \textit{sequential} and \textit{parallel}}
\vspace*{-0.25cm}

We compare the application of the two schemes on the CSR-V kernel. The \textit{sequential} kernel consumes more \textit{registers} and \textit{shared memory}, which limits the maximum number of warps that can be run on a SM to 24. In our application, CSR-S was limited to 24 warps/SM, for a bound of 64 warps/SM. This is reported in the column \textit{Theoretical Occupancy} of the following table. The low occupancy significantly decreases the performance. Concerning the \textit{global memory} access pattern, the column \textit{Load/Store efficiency} gives the ratio of requested memory transactions to the number of transactions performed, which reflects the degree to which memory accesses are coalesced (100\% efficiency) or not. For \textit{sequential} kernel, uncoalesced accesses cause the bandwidth loss and the performance degradation. The \textit{parallel} kernel makes the write accesses coalesced (100\% store efficiency). For the loads, it reaches only 47\% due to irregular accesses on the source vector. 

\vspace*{-0.25cm}


  \begin{center}
    \begin{tabular}{|l||c|c|c|c||c|}
      \hline
        & \footnotesize Registers & \footnotesize Shared Memory &  \footnotesize (Theoretical) & \footnotesize Load / Store & \footnotesize Timing\\
        & \footnotesize per thread & \footnotesize per SM & \footnotesize Occupancy & \footnotesize efficiency & \footnotesize in ms \\  
      \hline
      \hline
      \small \textit{Sequential} & \footnotesize 27 & \footnotesize 49152 & \footnotesize 35.1\% (37.5\%) & \footnotesize 7.5\% / 26\% & \footnotesize 141.1\\
      \bf \small \textit{Parallel} & \bf \footnotesize 21 & \bf  \footnotesize 15360 & \bf \footnotesize 70.3\% (100\%) & \bf \footnotesize 47.2\% / 100\% & \bf \footnotesize 41.4\\
	  \hline  
    \end{tabular}
  \end{center}

\vspace*{-0.25cm}

It is clear that the \textit{parallel} scheme is better suited to the context of large integers. We apply this scheme to other formats and compare their performance in the following subsection.

\vspace*{-0.25cm}
\subsection{Comparison of kernels CSR, COO, ELL and SLCOO}
\label{subsection: comparison kernels}
\vspace*{-0.25cm}

Due to the segmented reduction, the COO kernel performs more instructions and requires more registers. Thread divergence happens more often, because of the several branches that threads belonging to the same warp can take.

As far as the ELL kernel is concerned, the padded rows have the same length. This yields a good balancing across the warps and the threads (cf. \textit{Occupancy} and \textit{Branch divergence} in the following table). The column-major ordering makes this kernel reach the highest cache hit rate.  

The CSR-S kernel suffers from low efficiency of memory access compared to CSR-V. In fact, with the kernel CSR-S, the threads with the same warp work on several lines simultaneously, which makes their access to tables \texttt{id} and \texttt{data} not contiguous. The kernel CSR-V better satisfies the GPU architectural specificities.

\vspace*{-0.25cm}


  \begin{center}
    \begin{tabular}{|l||c|c|c|c|c||c|}
      \hline
        & \footnotesize Registers & \footnotesize Branch & \footnotesize (Theoretical) & \footnotesize Load / Store & \footnotesize Cache & \footnotesize Timing\\
        & \footnotesize per thread & \footnotesize divergence & \footnotesize Occupancy & \footnotesize efficiency & \footnotesize hit rate & \footnotesize in ms\\  
      \hline
      \hline
	  \footnotesize COO & \footnotesize 25 & \footnotesize 47.1\% & \footnotesize (66.7\%) 65.2\% & \footnotesize 34.3\%/37.8\% & \footnotesize 34.4\% & \footnotesize 88.9 \\
	  \hline
      \small CSR-S & \footnotesize 18 &  \footnotesize 28.1\% & \footnotesize (100\%) 71.8\% & \footnotesize 29.2\%/42.5\% & \footnotesize 35.4\% & \footnotesize 72.3 \\
	  \hline
      \bf \small CSR-V & \bf \footnotesize 21 & \bf \footnotesize 36.7\% & \bf \footnotesize (100\%) 70.3\% & \bf \footnotesize 47.2\%/100\% & \bf \footnotesize 35.4\% & \bf \footnotesize 41.4 \\
	  \hline
	  \footnotesize ELL-R & \footnotesize 18 & \footnotesize 44.1\% & \footnotesize (100\%) 71.8\%& \footnotesize 38.1\%/42.5\% & \footnotesize 40.1\% & \footnotesize 45.5 \\
	  \hline
    \end{tabular}
  \end{center}

\vspace*{-0.25cm}

The next table compares the CSR-V kernel with several SLCOO kernels for different slice sizes. We remark that increasing the slice size improves the cache hit rate, since accesses on the source vector are less irregular. However, by making the slices larger, we increase the usage of shared memory proportionally to the slice size, which limits the maximum number of blocks that can run concurrently. This limitation of the occupancy yields poor performance compared to the CSR-V kernel. Here, we consider an L1-oriented configuration (48kB L1, 16kB shared) of the on-chip memory. It is possible to move to a shared-oriented configuration (16kB L1, 48kB shared). This improves the occupancy, but degrades the cache hit rate, and finally does not improve the performances.

\vspace*{-0.25cm}


  \begin{center}
    \begin{tabular}{|l||c|c|c|c|c|c|}
      \hline
       & \footnotesize Shared Memory & \footnotesize Blocks & \footnotesize (Theoretical) & \footnotesize Cache & \footnotesize Timing \\
       & \footnotesize per Block & \footnotesize per SM & \footnotesize Occupancy  & \footnotesize hit rate & \footnotesize in ms \\  
      \hline
	  \hline
      \bf \footnotesize CSR-V & \bf \footnotesize 1920 & \bf \footnotesize 8 & \bf \footnotesize (100\%) 70.3\% & \bf \footnotesize 35.4\% & \bf \footnotesize 41.4 \\
      \hline
      \footnotesize SLCOO-2 & \footnotesize 3840 & \footnotesize 6 & \footnotesize (75\%) 68.4\% & \footnotesize 36.1\% & \footnotesize 46.9 \\
      \hline 	
      \footnotesize SLCOO-4 & \footnotesize 7680 & \footnotesize 4 & \footnotesize (50\%) 44.5\% & \footnotesize 36.9\% & \footnotesize 58.6 \\
      \hline
      \footnotesize SLCOO-8 & \footnotesize 15360 & \footnotesize 2 & \footnotesize (22.5\%) 22.3\% & \footnotesize 37.8\% & \footnotesize 89.9 \\
      \hline
    \end{tabular}
  \end{center}

\vspace*{-0.25cm}

Combinations of different formats have been tested. However they do not give better results. Splitting the matrix into column major blocks, or processing separately the first dense columns did not improve the performance either.    

For our matrix, the main bottleneck is memory access. In the CSR-V kernel, 72\% of the time is spent in reading data, 2\% in writing data and 26\% in computations.

\vspace*{-0.25cm}
\subsection{Comparison of RNS and Multi-precision arithmetics}
\vspace*{-0.25cm}

\label{RNS vs MP}
We also implemented the RNS and the multi-precision (MP) arithmetics on GPU. For the MP representation, to perform the reduction modulo $\ell$, we use a precomputed inverse of $\ell$ so as to divide by $\ell$ using a single multiplication. For the 217-bit prime order subgroup, choosing the largest representable integer $M=2^{256}-1$ is sufficient to accumulate a few number of SpMVs before reducing modulo $\ell$. In fact, the maximum row norm that we have (492) allows to do up to 4 iterative SpMVs before having to reduce.

For MP kernel, the reduction kernel takes only 0.37~ms, which corresponds to less than 0.1~ms per iteration. In RNS, we can accumulate 4 SpMVs before the reduction modulo~$\ell$ (cf. Table~\ref{Matrices}). The reduction kernel takes 1.6~ms (i.e., 0.4~ms per iteration).   

The idea behind the use of RNS rather than MP arithmetic is that RNS can significantly decrease data sharing between the threads and arithmetic operations required for the carry generation/propagation. The RNS kernel allows us to reach higher occupancy and better performance. The speed-up of RNS compared to multi-precision on the SpMV timing is around $15\%$.

\vspace*{-0.25cm}


\begin{center}
	\begin{tabular}{|l||c|c|c|c|c|c|}
      \hline
        & \footnotesize Registers & \footnotesize Shared Memory & \footnotesize Executed & \footnotesize (Theoretical) & \footnotesize Timing\\
        & \footnotesize per thread & \footnotesize per Block & \footnotesize Instructions & \footnotesize Occupancy & \footnotesize in ms\\  
      \hline
      \hline
      \footnotesize MP & \footnotesize 21 & \footnotesize 2880 & \footnotesize $6.1 \times 10^8$ & \footnotesize (83.3\%) 51.2\% & \footnotesize 46.6\\
      \bf \footnotesize RNS & \bf \footnotesize 18 & \bf \footnotesize 1920 & \bf \footnotesize $5.8 \times 10^{8}$ & \bf \footnotesize (100\%) 70.3\% & \bf \footnotesize 41.4\\
      \hline
    \end{tabular}    
\end{center}

\vspace*{-0.25cm}

\section{Improvements on CSR-V Kernel}
\label{optimizations}
To further improve the kernel performance, one should take into account the GPU architectural characteristics: the management of the memory accesses, the partitioning of the computations and the specificities of the problem considered.
	
\paragraph{\bf Texture caching}
Although our SpMV kernel suffers from irregular load accesses, a thread is likely to read from an address near the addresses that nearby threads (of the same group) read. For this reason, we bind on texture memory and replace reads with texture fetches. This improves the global memory efficiency and consequently the SpMV delay.     

\paragraph{\bf Reordering the non-zero coefficients of a row}
\label{Reordering non-zeros}
Since most of the coefficients of the matrix are $\pm1$, it seems promising to treat multiplications by these coefficients differently from other coefficients: additions and subtractions are less expensive than multiplications. All these separations result in code divergence, that we fix by reordering the non-zero coefficients in the matrix such that values of the same category ($+1, -1, >0, <0$) are contiguous. This decreases the branch divergence and decreases the total SpMV delay.  

\paragraph{\bf Compressing the values array \texttt{data}}
Since the majority of the coefficients are $\pm1$, after reordering the coefficients per row, we can replace the $\pm1$ coefficients by their occurrence count. This reduces the length of the values array \texttt{data} by more than 10 times, and so reduces the number of reads. 

\paragraph{\bf Improving warp balancing}
In the CSR-V kernel, each warp processes a single row. This requires launching a large number of warps. Consequently, there is a delay to schedule those launched warps. Instead, we propose that each warp iterates over a certain number of rows. To further increase the occupancy, we permute the rows such that each warp roughly gets the same work load.

\vspace*{-0.25cm}


	\begin{center}
    \begin{tabular}{|l||l|c|}
      \hline
      & \footnotesize Performance & \footnotesize Timing in ms \\
      & \footnotesize Effects & \footnotesize (speedup) \\
      \hline
      \hline
      {\footnotesize Texture caching} & {\footnotesize Global Load efficiency: 47.2\% $\rightarrow$ 84\%} & \footnotesize 32 (+30\%)\\
      \hline
      {\footnotesize Non-zeros reordering} & {\footnotesize Branch Divergence: 36.7\% $\rightarrow$ 12.9\%} & \footnotesize 30.5 (+5\%)\\
      \hline
      {\footnotesize Compressing \texttt{data}} & {\footnotesize Executed Instructions ($\times 10^8$): $5.8 $ $\rightarrow$ $5.72$} & \footnotesize 27.6 (+11\%)\\
      \hline
      {\small Multiple iterations} & {\footnotesize Occupancy: 70.3\% $\rightarrow$ 74.9\%} & \footnotesize 27.4 (+0.5\%)\\
      \hline
      {\small Rows permutation} & {\footnotesize Occupancy: 74.9\% $\rightarrow$ 81.8\%} & \footnotesize \textbf{27.1} (+1\%)\\
      \hline
    \end{tabular}
\end{center}

\vspace*{-0.25cm}

\section{Reference Software Implementation}

\label{soft}
For comparison purposes, we implemented SpMV on the three software instruction set architectures MMX, SSE and AVX, based on the RNS representation for the arithmetic and the CSR format for the storage of the matrix. We have not explored other formats that can be suitable for CPU. Probably blocked formats that better use the cache can further improve the performance on CPU. Unlike for GPU, processing separately the first dense columns accelerates the CPU SpMV of around 5\%.

We can report the computational throughput in terms of GFLOP/s, which we determine by dividing the number of required operations (twice the number of non-zero elements in the matrix $A$ multiplied by $2\times n$) by the running time. We will use this unit of measure for the throughput even for integer instructions.

The experiment was run on an Intel CPU i5-4570 (3.2~GHz) using 4 threads on 4 cores. The AVX2 implementation using integers is the fastest implementation and reaches the highest throughput. However, the fact that the number of moduli is a multiple of four entail overheads. When comparing the software performance with the GPU one, the fastest software implementation is 4 to 5 times slower than on one graphics processor. 

\vspace*{-0.25cm}


\begin{center}
    \begin{tabular}{|l||c|c|c|c|}
      \hline
        & \footnotesize Length of & \footnotesize Number of &  \footnotesize Timing & \footnotesize Throughput\\
	    & \footnotesize modulus & \footnotesize moduli ($n$) &  \footnotesize in ms & \footnotesize in GFLOP/s\\
      \hline
      \hline
      \footnotesize MMX (integer) & \footnotesize 64 & \footnotesize 5 & \footnotesize 306 & \footnotesize 5.3\\
      \footnotesize MMX (double-precision floats) & \footnotesize 52 & \footnotesize 6 & \footnotesize 351 & \footnotesize 3.7\\
      \footnotesize SSE2 (integer) & \footnotesize 63 & \footnotesize 6 & \footnotesize 154 & \footnotesize 12.1\\
      \footnotesize SSE2 (double-precision floats) & \footnotesize 52 & \footnotesize 6 & \footnotesize 176 & \footnotesize 10.6\\
      \bf \footnotesize AVX2 (integer) & \bf \footnotesize 63 & \bf \footnotesize 8 & \bf \footnotesize 117 & \bf \footnotesize 21.2\\
      \footnotesize AVX2 (double-precision floats) & \footnotesize 52 & \footnotesize 8 & \footnotesize 135 & \footnotesize 18.3\\
      \bf \footnotesize GPU (integer) & \bf  \footnotesize 64 & \bf \footnotesize 5 & \bf  \footnotesize 27 & \bf \footnotesize 57.6\\

      \hline
    \end{tabular}
\end{center}

\vspace*{-0.5cm}

\section{Conclusion}
\label{conclusion}
We have investigated different data structures to perform iterative SpMV for DLP matrices on GPUs. We have adapted the kernels for the context of large finite fields and added optimizations suitable to the sparsity and the specific computing model. The CSR-V kernel based on the \textit{parallel} scheme appears to be the most efficient one. The SLCOO poses for the sizes that we use some hardware difficulties that nullify its contribution on increasing the cache hit rate. Future GPUs may enhance the performance. We have shown that using RNS for finite field arithmetic provides a considerable degree of independence, which can be exploited by massively parallel hardware. This implementation contributed to solving the discrete logarithm problem in GF($2^{619}$) and GF($2^{809}$) (See Appendix~\ref{Appendix::Linear Algebra} and~\cite{FFS809,JEL14} for further details). 
%

\vspace*{-0.25cm}

\bibliographystyle{plain}
\bibliography{spmv}

\vspace*{-0.75cm}

\appendix
\section{Formats and GPU Kernels of SpMV}
\label{Appendix:: kernels}

\begin{figure}[h]

\vspace*{-1.25cm}

\centering

\begin{tabular}{C{.31\textwidth}C{.69\textwidth}}

\begin{subfigure}[]{0.30\textwidth}
	\centering
	\begin{tikzpicture}[scale=0.7]
		\draw [transparent](-2,-2) rectangle (2,2);
		\tikzstyle{every right delimiter}=[xshift=-1
		.5ex]
		\tikzstyle{every left delimiter}=[xshift=1.5ex]
		\pgflowlevelsynccm
		\matrix [matrix of math nodes,left delimiter=(,right delimiter=)] at (0,0)
		{
		0 & \color{red} a_{01} & 0 & \color{red} a_{03} & 0 & 0 \\
		0 & \color{blue} a_{11} & 0 & 0 & \color{blue} a_{14} & \color{blue} a_{15} \\
		\color{orange} a_{20} & 0 & \color{orange} a_{22} & \color{orange} a_{23} & 0 & 0 \\
		0 & a_{31} & 0 & 0 & a_{34} & 0 \\
		0 & a_{41} & a_{42} & 0 & 0 & a_{45} \\
		0 & 0 & a_{52} & 0 & 0 & a_{55} \\
		};
	\end{tikzpicture}
	\caption{Sparse matrix $A$}
\end{subfigure}

& 

\begin{subfigure}[]{0.68\textwidth}

	\begin{tikzpicture}[scale=0.7]
		\draw [transparent](-5.5,0.75) rectangle (2,-2.5);
			\pgflowlevelsynccm
			\node at (-4.875,-0.875){\texttt{data = }};
			\tikzstyle{every right delimiter}=[xshift=-1.5ex]
			\tikzstyle{every left delimiter}=[xshift=1.5ex]
			\matrix [matrix of math nodes,left delimiter={[},right delimiter={]}] at (-3.075,-0.875)
			{
			\color{red} a_{01} & \color{red} a_{03} & * \\
			\color{blue} a_{11} & \color{blue} a_{14} & \color{blue} a_{15} \\
			\color{orange} a_{20} & \color{orange} a_{22} & \color{orange} a_{23}   \\
			a_{31} & a_{34} & * \\
			a_{41} & a_{42} & a_{45} \\
			a_{52} & a_{55} & * \\
			};
			
			\node at (-1.3,-0.875){\texttt{id = }};
			\tikzstyle{every right delimiter}=[xshift=-1.5ex]
			\tikzstyle{every left delimiter}=[xshift=1.5ex]
			\matrix [matrix of math nodes,style={nodes={minimum width=1.5em}},left delimiter={[},right delimiter={]}] at (0.025,-0.875)
			{
			\color{red} 1 & \color{red} 3 & * \\
			\color{blue}1 & \color{blue} 4 & \color{blue}5 \\
			\color{orange} 0 & \color{orange} 2 & \color{orange} 3 \\
			1 & 4 & * \\
			1 & 2 & 5 \\
			2 & 5 & * \\
			};	

			\node at (1.55,-0.875){\texttt{len = }};
			\tikzstyle{every right delimiter}=[xshift=-1.5ex]
			\tikzstyle{every left delimiter}=[xshift=1.5ex]
			\matrix [matrix of math nodes,style={nodes={minimum width=1.2em}},left delimiter={[},right delimiter={]}] at (3.425,-0.875)
			{
			\color{red} 2 & \color{blue} 3 & \color{orange} 3 & 2 & 3 & 2 \\
			};	

	\end{tikzpicture}
	\caption{ELL-R representation}	
	
\end{subfigure}

\\
&
\begin{subfigure}[]{0.7\textwidth}

	\begin{tikzpicture}[scale=0.7]
		\draw [transparent](-5.375,1.25) rectangle (5.3875,-2.5);
			\pgflowlevelsynccm
			\node at (-4,0.75){\texttt{data = }};
			\tikzstyle{every right delimiter}=[xshift=-1.5ex]
			\tikzstyle{every left delimiter}=[xshift=1.5ex]
			\matrix [matrix of math nodes,left delimiter={[},right delimiter={]}] at (0.8,0.75)
			{
			\color{red} a_{01} & \color{red} a_{11} & \color{red} a_{03} & \color{red} a_{14} & \color{red} a_{15} & \color{blue} a_{20} & \color{blue} a_{31} & \color{blue} a_{22} & \color{blue} a_{23} & \color{blue} a_{34} &\dots\\
			};
			
			\node at (-4.2,-0.25){\texttt{row\_id = }};
			\tikzstyle{every right delimiter}=[xshift=-1.5ex]
			\tikzstyle{every left delimiter}=[xshift=1.5ex]
			\matrix [matrix of math nodes,style={nodes={minimum width=2em}},left delimiter={[},right delimiter={]}] at (0.8,-0.25)
			{
			\color{red} 0 & \color{red} 1 & \color{red} 0 & \color{red} 1 & \color{red} 1 & \color{blue} 2 & \color{blue} 3 & \color{blue} 2 & \color{blue} 2 & \color{blue} 3 & \dots\\
			};

			\node at (-4.2,-1.25){\texttt{col\_id = }};
			\tikzstyle{every right delimiter}=[xshift=-1.5ex]
			\tikzstyle{every left delimiter}=[xshift=1.5ex]
			\matrix [matrix of math nodes,style={nodes={minimum width=2em}},left delimiter={[},right delimiter={]}] at (0.8,-1.25)
			{
			\color{red} 1 & \color{red} 1 & \color{red} 3 & \color{red} 4 & \color{red} 5 & \color{blue} 0 & \color{blue} 1 & \color{blue} 2 & \color{blue} 3 & \color{blue} 4 & \dots\\
			};
			
			\node at (-4.4,-2.25){\texttt{ptrSlice = }};
			\tikzstyle{every right delimiter}=[xshift=-1.5ex]
			\tikzstyle{every left delimiter}=[xshift=1.5ex]
			\matrix [matrix of math nodes,style={nodes={minimum width=2em}},left delimiter={[},right delimiter={]}] at(-1.65,-2.25)
			{
			\color{red} 0 & \color{blue} 5 & 10 &  &  &  &  &  & \dots\\
			};	
	\end{tikzpicture}
	\caption{SLCOO-2 representation}	
	
\end{subfigure}

\end{tabular}

\vspace*{-1.25cm}

\end{figure}

\begin{algorithm}[H]
 \SetKwInOut{Input}{Inputs}\SetKwInOut{Output}{Output}
 \SetKwFor{For}{For}{do}{endfor}
 \SetKwFor{While}{While}{do}{endwhile}
 \SetKwFor{If}{If}{then}{else}
 \Input{\small \texttt{data}: array of $K \times N$ elements of $K$, \texttt{id}: array of $K \times N$ positive integers, \\
 \texttt{len}: array of $N$ positive integers and $u$: vector of $N$ elements of $K$.
 }
 \Output{\small $v$: vector of $N$ elements of $K$.}
 \BlankLine
 \small 
 \texttt{sum} $\leftarrow 0$\;
  \For{$j\leftarrow 0$ \KwTo \texttt{\upshape len}\textsubscript{$i$}}{
  		\texttt{sum} $\leftarrow$ \texttt{addmul(sum,data}\textsubscript{$N\times j+$\texttt{\upshape row}}\texttt{,}$u$\textsubscript{\texttt{id}\textsubscript{$N\times j+$\texttt{\upshape row}}}\texttt{)}\;
  }
 $v_i$ $\leftarrow$ \texttt{sum}\; 
 \caption{\small ELL-R for row $i$ executed by one thread}
\end{algorithm}

\begin{algorithm}[H]
 \SetKwInOut{Input}{Inputs}\SetKwInOut{Output}{Output}
 \SetKwFor{For}{For}{do}{endfor}
 \SetKwFor{While}{While}{do}{endwhile}
 \SetKwFor{If}{If}{then}{else}
 \Input{\small \texttt{data}: array of $n_{NZ}$ elements of $K$,
 \texttt{row\_id}, \texttt{col\_id}: arrays of $n_{NZ}$ positive integers,
 \texttt{ptrSlice}: array of $N$ positive integers and
 $u$: vector of $N$ elements of $K$.
 }
 \Output{\small $v$: vector of $N$ elements of $K$.}
 \small 
 \BlankLine
 Declare array \texttt{sum} $\leftarrow \{0\}$\tcp*[r]{array of $\sigma$ elements of $K$}
 $i\leftarrow$ \texttt{ptrSlice}\textsubscript{$i$} $+$ \texttt{tid}\tcp*[r]{position of beginning for each thread}
  \While{$j<$ \texttt{\upshape ptrSlice\textsubscript{$i+1$}}}{
  	\texttt{sum}\textsubscript{\texttt{row\_id}\textsubscript{$j$}$\bmod \sigma$} $\leftarrow$ \texttt{addmul(}\texttt{sum}\textsubscript{\texttt{row\_id}\textsubscript{$j$}$\bmod$ $\sigma$}\texttt{,}\texttt{data}\textsubscript{$j$}\texttt{,} $u$\textsubscript{\texttt{col\_id}\textsubscript{$j$}$\bmod$ $\sigma$}\texttt{)}\;
   	$j \leftarrow j + 32$\;
   }
   \texttt{reduction\_slcoo(sum,tid)}\tcp*[r]{reduction in \textit{shared memory}}
  \If(\tcp*[f]{first thread of warp writes in \textit{global memory}}){\texttt{\upshape tid}$=0$}{
  	\For{$j\leftarrow 0$ \KwTo $\sigma$}{
		$v$\textsubscript{\texttt{$i\times \sigma + j$}} $\leftarrow$ \texttt{sum}\textsubscript{$j$}\;
  	}
  } 
 \caption{\small SLCOO-$\sigma$ for slice $i$ executed by thread of index \texttt{tid} in its warp}
\end{algorithm}

\section{Resolution of Linear Algebra of the Function Field Sieve}

\label{Appendix::Linear Algebra}
The linear algebra step consists of solving the system  $Aw=0$, where $A$ is the matrix produced by the filtering step of the FFS algorithm. $A$ is singular and square. Finding a vector of the kernel of the matrix is generally sufficient for the FFS algorithm. 

The simple Wiedemann algorithm~\cite{WIED86} which resolves such a system, is composed of three steps: 
\begin{itemize}
\item \textit{Scalar products}: It consists on the computation of a sequence of scalars $a_i = \,^txA^i y$, where $0 \leq i \leq 2N$, and $x$ and $y$ are random vectors in $(\mathbb{Z}/\ell\mathbb{Z})^N$. We take $x$ in the canonical basis, so that instead of performing a full dot product between $^tx$ and $A^i y$, we just store the element of $A^i y$ that corresponds to the  non-zero coordinate of $x$.
\item \textit{Linear generator}: Using the Berlekamp-Massey algorithm, this step computes a linear generator of the $a_i$'s. The output $F$ is a polynomial whose coefficients lie in $\mathbb{Z}/\ell\mathbb{Z}$, and whose degree is very close to $N$. 
\item \textit{Evaluation}: The last step computes $\sum_{i=0}^{deg(F)}{A^iF_iy}$, where $F_i$ is the $i^{th}$ coefficient of $F$. The result is with high probability a non-zero vector of the kernel of $A$.
\end{itemize} 

The Block Wiedemann algorithm~\cite{KALT95} proposes to use $m$ random vectors for $x$ and $n$ random vectors for $y$. The sequence of scalars is thus replaced by a sequence of $m \times n$ matrices and the numbers of iterations of the first and third steps become $(N/n + N/m)$ and $N/n$, respectively. The $n$ subsequences can be computed independently and in parallel. So, the block Wiedemann method allows to distribute the computation without an additional overhead~\cite{THOM02}. 

\subsection{Linear Algebra of FFS for GF($2^{619}$)}
The matrix has 650k rows and columns. The prime $\ell$ is 217 bits.
The computation was completed using the simple Wiedemann algorithm on a single NVIDIA GeForce GTX 580. The overall computation needed 16 GPU hours and 1 CPU hour. 

\subsection{Linear Algebra of FFS for GF($2^{809}$)}
The matrix has 3.6M rows and columns. The prime $\ell$ is 202 bits.
We run a Block Wiedemann on a cluster of 8 GPUs. We used 4 distinct nodes, each equipped with two NVIDIA Tesla M2050 graphics processors, and ran the Block Wiedemann algorithm with blocking parameters $m=8$ and $n=4$. The overall computation required 4.4 days in parallel on the 4 nodes.

\medskip 
These two computations were part of record-sized discrete logarithm computations in a prime-degree extension field~\cite{FFS809,JEL14}.

\end{document}